\documentclass[twocolumn,showpacs,amsfonts,amsmath,assymb,eufrak,prl]{revtex4-1}
\usepackage{epsfig}
\usepackage{color}
\usepackage{bm}
\usepackage{braket}
\usepackage{graphicx}

\newcommand{\eq}[1]{\begin{equation} #1 \end{equation}}
\newcommand{\eqa}[2]{\begin{equation} #1 \label{#2} \end{equation}}
\newcommand{\balign}[1]{\begin{align} #1 \end{align}}

\newcommand{\figin}[4]
{\begin{figure}[tb]
\centering
\includegraphics[width= #1]{#2}
\caption{#3}
\label{f:#4}
\end{figure}}

\newcommand{\todayd}{\the\year/\the\month/\the\day}

\newcommand{\bib}{\bibitem}

\newcommand{\up}{\uparrow}
\newcommand{\down}{\downarrow}

\newcommand{\lb}{\label}
\newcommand{\nt}{\notag}
\newcommand{\Tr}{\mathrm{Tr}}

\newcommand{\bel}{\begin{easylist}}
\newcommand{\eel}{\end{easylist}}

\newcommand{\fref}[1]{Fig.~\ref{f:#1}}

\def \({\left(}
\def \){\right)}
\def \[{\left[}
\def \]{\right]}
\newcommand{\la}{\left\langle}
\newcommand{\ra}{\right\rangle}
\newcommand{\abs}[1]{\left|#1\right|}

\newcommand{\sumtwo}[2]%
{\mathop{\sum_{#1}}_{#2}}
\newcommand{\sumthree}[3]%
{\mathop{\mathop{\sum_{#1}}_{#2}}_{#3}}
\newcommand{\sumfour}[4]%
{\mathop{\mathop{\mathop{\sum_{#1}}_{#2}}_{#3}}_{#4}} 
\newcommand{\prodtwo}[2]%
{\mathop{\prod_{#1}}_{#2}}
\newcommand{\mintwo}[2]%
{\mathop{\min_{#1}}_{#2}}
\newcommand{\maxtwo}[2]%
{\mathop{\max_{#1}}_{#2}}
\newcommand{\maxthree}[3]%
{\mathop{\mathop{\max_{#1}}_{#2}}_{#3}}
\newcommand{\limtwo}[2]%
{\mathop{\lim_{#1}}_{#2}}
\newcommand{\suptwo}[2]%
{\mathop{\sup_{#1}}_{#2}}
\newcommand{\supthree}[3]%
{\mathop{\mathop{\sup_{#1}}_{#2}}_{#3}}
\newcommand{\supfour}[4]%
{\mathop{\mathop{\mathop{\sup_{#1}}_{#2}}_{#3}}_{#4}} 
\newcommand{\inftwo}[2]%
{\mathop{\inf_{#1}}_{#2}}
\newcommand{\infthree}[3]%
{\mathop{\mathop{\inf_{#1}}_{#2}}_{#3}}
\newcommand{\inffour}[4]%
{\mathop{\mathop{\mathop{\inf_{#1}}_{#2}}_{#3}}_{#4}} 

\newcommand\calB{{\cal B}}

\newcommand\calL{{\cal L}}

\newcommand{\Di}{\mathit{\Delta}}

\newcommand{\para}[1]{{\em #1}\/.---}

\def\rnum#1{\resizebox{0.5em}{\height}{\expandafter{\romannumeral #1}}}
\def\Rnum#1{\resizebox{0.5em}{\height}{\uppercase\expandafter{\romannumeral #1}}}

\newcommand{\HS}{H_{\rm S}}

\makeatletter
\renewcommand{\@cite}[1]{\textsuperscript{#1)}}
\makeatother

\begin{document}

\preprint{APS/123-QED}

\newcommand{\titlename}{Speed limit for open systems coupled to general environments}

\preprint{APS/123-QED}

\title{\titlename}

\author{Naoto Shiraishi$^{1}$ and Keiji Saito$^{2}$ }
\affiliation{$^{1}$Department of Physics, Gakushuin University, 1-5-1 Mejiro, Toshima-ku, Tokyo, Japan} 

\affiliation{$^{2}$Department of Physics, Keio university, Hiyoshi 3-14-1, Kohoku-ku, Yokohama, Japan}%

\date{\today}

\begin{abstract}
 In this study, we investigate the bound on the speed of state transformation in the quantum and classical systems that are coupled to general environment with arbitrary coupling interactions. We show that a Mandelstam-Tamm type speed limit exists and energy fluctuation still plays a crucial role in this speed limit inequality for open quantum systems. The energy fluctuation of the target system in addition to the coupling to the environment is key in the inequality. We also present the classical version of the speed limit for open systems. As potential applications of the proposed speed limit expression, we discuss the fundamental limitation of the state change in quantum cyclic engines and the equilibriation time required for the thermalization phenomena of isolated quantum systems.
\end{abstract}

\maketitle


\para{Introduction}
The rate at which a given initial state can be transformed into a desired final state has received significant research attention in the field of theoretical physics. Moreover, questions pertaining to the quantities limiting the speed of state change have been thoroughly investigated. One of the answers is obtained via the celebrated Mandelstam-Tamm (MT) quantum speed limit for isolated quantum systems~\cite{MT}, which is expressed by the following inequality: 
\eqa{
\frac{\calL(\rho(\tau), \rho(0))}{\la \delta E\ra_\tau /\hbar}\leq \tau \, .
}{MT}
Here, $\la \delta E\ra_\tau$ is the time-average of the energy fluctuation during the operation time $\tau$, and $\calL (\rho, \rho')$ is the Bures angle, which is a type of distance, between the two states (i.e., two density matrices) $\rho$ and $\rho'$~\cite{SM}. This inequality clearly indicates that the amount of energy fluctuation restricts the speed of state change. Thereafter, Margolus and Levitin proposed another type of speed limit, where energy itself, and not energy fluctuation, bounds the quick state transformation~\cite{ML}. These relations originated from the uncertainty relation between time and energy; hence, the operation time is inevitably accompanied by quantities on energy~\cite{MT, Fle, AA, Pre, ML, GLM, JK,DC, Funo}.

During experiments, the environmental effect is unavoidable. A typical example is solid-state quantum devices, which are frequently disturbed by thermal as well as nonthermal environments~\cite{Brebook}. Therefore, it is fundamentally important to understand the effects of the environment on the speed of state transformation, and several studies have focused on this aspect. When the dynamics of a system is well described by an effective equation, such as the Lindblad equation, mathematically compact expressions on the speed limit can be derived~\cite{Cam, DL, Zha,Pir, FSS}. Nonunitary dynamics arising from the environment has also been discussed, from the perspective of quantum information theory~\cite{Tad}. It is also intriguing that the speed of state transformation can be discussed in terms of the entropy production rate induced due to the state change and the thermal environment, using stochastic thermodynamics frameworks \cite{Sekimoto-Sasa, Aurell,BSS,SST,ST,SFS,SS}

It should be noted that in several realistic situations, we can see a wide variety of environments, coupling forms and strengths between the system and environment. Not only the nearest interaction, but also Lenard-Jones interaction, and the long-range interaction (e.g. the dipole interaction), are commonly observed in various materials. Remarkably, recent experiments have enabled us to control the range of long-range interaction~\cite{long1,long2,long3,long4}. The circuit quantum electrodynamics is known to have strong coupling strength between the two-level system and the electromagnetic environment in the circuit. Generally, the environment does not necessarily act as a thermal reservoir, but as nonthermal quantum noise, and the dynamics of the target system can be significantly complex, making it difficult to describe using the compact form of an equation of motion. 
Based on this background, we take the question one-step further, i.e., is there any general and physically meaningful bound on the speed of state transformation that holds irrespective of the type of environment, coupling form, and coupling strength? Given the prevalence of several relevant experimental situations, it is necessary to address this question for an in-depth understanding of the speed of state transformation observed in nature.

In this Letter, we aim to clarify the limitation of state transformation by deriving the speed limit expression for systems attached to the general environment through arbitrary coupling. We establish that the energy fluctuation still plays a central role in limiting the speed of state transformation. We show that the energy fluctuation of the coupling component as well as the system Hamiltonian is crucial. To derive the speed limit inequality, we apply the decomposition on the time evolution operator, separating it into two contributions: one coming from the system plus coupling, and the other coming from the environment. By extending this idea to the classical case together with this new method \cite{Campo}, we also derive the classical version of the speed limit for Hamiltonian dynamics. The speed limit expression does not only show the fundamental mechanism of state transformation, but also has several potential applications. As simple applications, we discuss the microscale quantum engines and the equilibration time in thermalization phenomena of isolated quantum many-body systems.

\para{Speed limit for open quantum systems}
We consider a quantum system (S) interacting with the external environment(E) via coupling (SE) with an arbitrary interacting form. The total Hamiltonian is described as
\begin{align}
  H_{\rm tot} &= H_{\rm S} + H_{\rm SE} + H_{\rm E} \, , \label{ham}
\end{align}
where $H_{\rm S}$ and $H_{\rm E}$ are the Hamiltonian of the system and that of the environment, respectively. The Hamiltonian $H_{\rm SE}$ is a coupling Hamiltonian between the system and the environment. We emphasize that $H_{\rm SE}$ is arbitrary, and its norm can be the same order as the system's Hamiltonian. Although we do not write the time-dependence explicitly for the sake of notational simplicity, each Hamiltonian can be time-dependent. 

Let $\rho_{\rm tot}(t)$ and $\rho_{\rm S}(t)$ the density matrix for the total system and the target system of interest, respectively. In the time evolution driven by $H_{\rm tot}$, the total density matrix changes in time, and the system's reduced density matrix is obtained as $\rho_{\rm S}(t) = \Tr_{\rm E}[\rho_{\rm tot}(t)]$, where $\Tr_{\rm E}$ implies the partial trace of the environment. Then, we consider a speed limit only for the system by analyzing the change in the system's density matrix. Let $\calL(\rho, \rho') $ be the Bures angle, which quantifies the distance between two states $\rho$ and $\rho'$, defined as \cite{NC}
\begin{align}
  \calL(\rho, \rho') :=\arccos \Tr[\sqrt{\rho^{1/2}\rho' \rho^{1/2}}] \, .
\end{align}
In addition, we define the system plus the interaction Hamiltonian denoted by $\tilde{H}_{\rm S}$
and the energy variance for this Hamiltonian at time $t$ denoted by $\delta {E_{\rm S}} (t)$:
\begin{align}
  \tilde{H}_{\rm S}  &:=H_{\rm S} + H_{\rm SE} \, , \\
  \delta{E_{\rm S}} (t)  & := \sqrt{ \Tr [ \tilde{H}_{\rm S}^2 {\rho}_{\rm tot}(t)]- \Tr [ \tilde{H}_{\rm S} {\rho}_{\rm tot}(t)]^2 } \, .
\end{align}
The Hamiltonian $\tilde{H}_{\rm S}$ is a collection of all terms containing the system's operators.
Then, we derive the following quantum speed limit for general open systems:
\eqa{
\frac{\calL(\rho_{\rm S}(\tau), \rho_{\rm S}(0))}{\la \delta E_{\rm S}\ra_\tau /\hbar}\leq \tau \, ,
}{main-q}
where $\la \delta {E}_{\rm S}\ra_\tau$ is the time-average on the energy fluctuation, i.e., $\la \delta {E}_S\ra_\tau:=(1/\tau) \int_0^\tau dt \delta E_S(t)$. This is the first main result of this Letter. The proof for this relation is provided at the end of this Letter. 

We shall clarify the physical meaning behind this inequality. We note that the bound on the operation time in (\ref{main-q}) is characterized not only by the system Hamiltonian in the energy fluctuation but also by the coupling component. This combination is essential for the correct estimation of the bound, which can be understood by comparing it with the direct application of the MT bound (\ref{MT}). 
Firstly, if we employed the original MT bound (\ref{MT}), setting the energy fluctuation only by the system Hamiltonian, then we immediately realize that this inequality does not hold.
A simple counterexample is the case where the system Hamiltonian is constant, while the coupling Hamiltonian exists, in which the energy fluctuation of the system Hamiltonian is exactly zero. 
Next, suppose that we employed the original MT bound (\ref{MT}) by setting the energy fluctuation for the total Hamiltonian and applying the monotonicity of the Bures angle $\calL (\rho_{\rm S} (\tau), \rho_{\rm S} (0))\le \calL (\rho_{\rm tot} (\tau), \rho_{\rm tot} (0))$. 
Then, the original MT bound can be written in terms of the state change only for the system: 
$\calL (\rho_{\rm S} (\tau), \rho_{\rm S} (0))/ (\la \delta {E}_{\rm tot}\ra_\tau/ \hbar ) \le \tau$, where $\la \delta {E}_{\rm tot}\ra_\tau$ is the time-average of energy fluctuation for the total Hamiltonian. 
This bound is a correct inequality but it is very different from the relation (\ref{main-q}), especially when the size of the environment is very large. In the case of an infinite environment size, which is a typical setting for the thermal reservoir, the total energy fluctuation $\la \delta {E}_{\rm tot}\ra_\tau$ diverges, leading to a trivial bound
$ 0 \le \tau $. On the other hand, the bound in (\ref{main-q}) gives a finite value even in this case. In some cases, the original MT bound is not available, while the relation (\ref{main-q}) works well.

We should also note that the energetic effects of the environment can be achieved only via the coupling Hamiltonian. Without the finite coupling Hamiltonian, the environment cannot affect the Bures angle inside the system, even if quantum nonlocal effects, such as quantum entanglement, exist between the system and environment. The effect of the environment reaches the system with finite time, provided the energy fluctuation, including the coupling Hamiltonian, is finite. This argument also supports the structure including the coupling Hamiltonian in the speed limit expression (\ref{main-q}).

\para{Speed limit for open classical Hamiltonian systems}
We extend the speed limit (\ref{main-q}) to classical Hamiltonian systems.
Although several speed limits for classical Hamiltonian systems have been proposed~\cite{Def, Campo, ohzeki}, we employ the technique given in Ref.~\cite{Campo}. Consider a classical version of the total Hamiltonian (\ref{ham}), and the distribution function in the phase space at time $t$, $\rho_{\rm tot} (t)$. The dynamics of $\rho_{\rm tot} (t)$ is given by the Liouville equation $(\partial/\partial t) \rho_{\rm tot} (t) =\{ H_{\rm tot}, \rho_{\rm tot} (t) \}$, where $\{ A, B\} :=\sum_{\alpha=x,y,z}\sum_i (\partial A / \partial  q_{i,\alpha} )(\partial B / \partial  p_{i,\alpha} )- (\partial A / \partial  p_{i,\alpha} )(\partial B / \partial  q_{i,\alpha} )$ is the Poisson bracket, and $q_{i,\alpha}$ and $p_{i,\alpha}$ are, respectively, the position and momentum variables of the $i$-th particle in the $\alpha$-th direction.
We employ the classical Bures angle (Bhattacharyya angle) as the distance between two classical distribution functions:
\begin{align}
  \begin{split}
\calL_B(\rho, \rho')&:=\arccos \calB , \\
\calB  &:=\int d \Gamma \sqrt{ \rho } \sqrt{ \rho^{\prime}} \, , 
\end{split}
\end{align}
where $d\Gamma =\prod_{\alpha}\prod_i dp_{i,\alpha}dq_{i,\alpha}$, and $\calB$ is the Bhattacharyya coefficient, which serves as a classical counterpart of the fidelity.

The distribution function for the system is defined as $\rho_S:=\int d\Gamma_E \rho_{\rm tot}$, where $\int d\Gamma_E$ is integration with respect to the degree of freedom in the environment. We then derive the speed limit for classical open Hamiltonian systems:
\eqa{
  \frac{(2/ \pi)\calL_B(\rho_S(\tau), \rho_S(0))}{\la \sqrt{\int d\Gamma \, \{ \tilde{H}_{\rm S} , \sqrt{\rho_{\rm tot}}\}^2} \ra _\tau  }\leq \tau ,
}{main-c}
where $\langle ... \rangle_{\tau}$ implies the average over time. This classical expression is our second main result. Similar to the quantum version (\ref{main-q}), the bound contains the Hamiltonian $\tilde{H}_{\rm S}$. We present the details of the derivation in the Supplemental Material \cite{SM}.

\para{Solvable toy example}
We here demonstrate the speed limit in the open quantum system (\ref{main-q}) with a solvable model. 
We use a globally coupled spin-1/2 system that demonstrates the validity of (\ref{main-q}), even for long-range coupling with arbitrary strength. 
We regard a single spin as a principal system and other $N$ spins attached to this single spin as the environment. We focus on the state transformation of the single spin. 
We set the Hamiltonians as
\begin{align}
\begin{split}
H_{\rm S}&= h S_1^{z} \, , ~~
H_{\rm SE} = J\sum_{i=2}^{N+1} {\bm S}_1 \cdot {\bm S}_{i}\, , \\
H_{\rm E} &= h \sum_{i=2}^{N+1} S_{i}^z  + J \sum_{i=2}^{N+1}\sum_{j > i}^{N+1} {\bm S}_i \cdot {\bm S}_{j} \, ,
\end{split}
\end{align}
where ${\bm S}_{i}$ is a vector of spin-1/2 operator at the site $i$, i.e., ${\bm S}_i =(S_i^x , S_i^y, S_i^z)$. Let $\ket{\up}_i$ ($\ket{\down}_i$) be the eigenstate of the z-component of the spin at the site $i$, satisfying $S^z_i \ket{\up}_i=\frac12 \ket{\up}_i$ ($S^z_i \ket{\down}_i=-\frac12 \ket{\down}_i$). 
For simplicity, we set the Hamiltonian as a static one. 
We set the initial state $\rho (0) =|\psi (0) \rangle \langle \psi (0) |$ with $|\psi (0) \rangle =|\uparrow \rangle_1 \otimes | \downarrow \rangle_2 \cdots |\downarrow \rangle_{N+1}$. Using the conservation law on the total magnetization, it is straightforward to calculate the time-evolution of the state. The Bures angle for the one spin system during the time $\tau$ and the energy fluctuation for the Hamiltonian $\tilde{H}_{\rm S}$ at time $t$ $(0\leq t\leq \tau$) is computed as
\begin{align}
{\cal L} (\rho_{\rm S} (\tau) , \rho_{\rm S} (0) ) =& \arccos \left[ { \sqrt{(N-1)^2 + 4 N \cos^2 \left( \frac{JN \tau}{4\hbar} \right)}\over N+1}\right] \, , \nonumber \\
\left[ \delta E_{\rm S}(t) \right]^2 =&
N \Bigl\{ \left[ {J \over 2}- { 2 h (N-1) \over (N+1)^2 }\sin^2 \left( \frac{JNt}{4\hbar} \right) \right]^2 \nonumber \\
& \hspace{12pt} + \left[{h\over N+1} \sin \left( \frac{JNt}{4\hbar}\right)\right]^2 
\Bigr\} \, . \nonumber
\end{align}
For $N=1$, these expressions are reduced to ${\cal L} = J \tau /4\hbar$ and $\delta E_{\rm S} (t) = \left[ (J/2)^2 + ((h/2) \sin (J t /4\hbar))^2\right]^{1/2}$. These clearly satisfy the relation (\ref{main-q}). For a sufficiently small $\tau$, the left hand side in (\ref{main-q}) is approximately equal to $N\tau/(N+1)$ for any $N$, which is tight for a large $N$. In \fref{figex}, we plot the left-hand side of \eqref{main-q} versus $\tau$ for several sizes of $N$ with the fixed parameter $h/J=1$, which clearly indicates the validity of \eqref{main-q} at all times.

\figin{7.5cm}{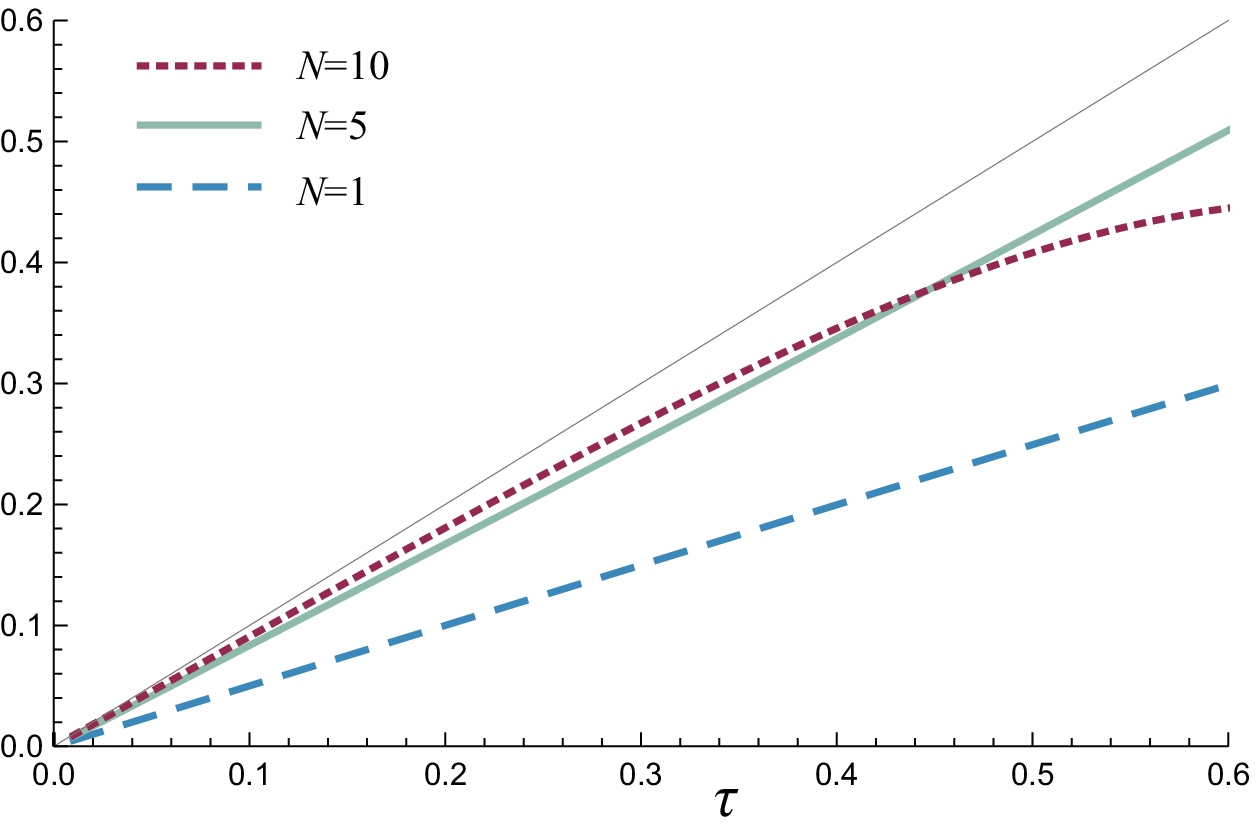}{
Speed limit bounds (left-hand side of \eqref{main-q}) versus actual time $\tau$ for different sizes of $N$ with the parameter $h/J=1$. The unit of time is $\hbar/J$.
The thin gray line represents $y=\tau$.
All plots are under the gray line, as shown in \eqref{main-q}.
}{figex}

\para{Application to cyclic quantum engines}
We show that the relation (\ref{main-q}) has important implications to several physical research fields. As the first application, we consider cyclic quantum engines surrounded by large environments; examples include microscale quantum heat engines \cite{review} and parametric electron pumping \cite{reviewdriven,thouless,brouwer}, both of which have been realized in recent experiments\cite{atomexp,nvcenter,switkes,takahashi}.

In theoretical description of heat engines driven by two different temperatures, we should control the amplitude of the interactions between the system and the heat baths to switch the temperatures. By changing temperatures in this manner and controlling the system's parameters, we extract quantum work through a cycle process with the period $\tau_{\rm c}$. The cyclic property requires that the initial and the final states of the system be identical, $\rho_{\rm S} (t+\tau_{c})=\rho_{\rm S} (t)$. Then, the relation (\ref{main-q}) translates to the following relation \cite{comment-derivation}
\begin{align}
   \max_{\rho _{\rm S} (t) , \rho_{\rm S} (t')} {{\cal L} (\rho_{\rm S} (t) , \rho_{\rm S} (t') )} 
  \le  { \tau_{\rm c}   \delta E_{\rm S, max} \over 2 \hbar }
  \le { \tau_{\rm c}  \| \tilde{H}_{\rm S} \|  \over 2 \hbar }\, ,
\end{align}
where $\delta E_{\rm S, max}$ is the maximum value of $\delta E_{\rm S} (t)$ during the cycle, and $\| ... \|$ is a spectral norm. The maximum in the left-hand side is taken over all time duration $0\leq t,t'\leq \tau$. This relation must be satisfied in any type of cyclic engines, including microscopic heat engines and parametric electron pumping. This relation indicates a fundamental limitation of the speed of engines accompanying state change, which should be significant as the size of target system becomes small and as a result, the quantum effect becomes important.

\para{Application to thermalization in isolated quantum systems}
As a second application, we consider the thermalization phenomena of isolated large quantum systems. We make a rough estimation of the {\it equilibration time} from a certain initial pure state~\cite{Neu, Rig08, Rei08,LPSW,SF12, MRA, Pal}. To this end, we need to restrict our attention to macroscopic observables~\cite{Neu, Rig08, Rei08, SF12} or small subsystems~\cite{LPSW, MRA, Pal}. We here consider the latter case, and regard the subsystem as a target system and the other component as the environment. We assume that the state of the total system equilibrates (i.e., relaxes to a certain macroscopic state) and the reduced density matrix of the subsystem reaches a steady state within a finite time.

Although the thermalization phenomena is widespread, estimating the equilibration time is considerably difficult. Calculations based on random Hamiltonians or typicality argument have been successfully performed, while they exhibit extremely fast equilibration time~\cite{GHH, Rei16}. Analyses based on realistic Hamiltonian are highly complicated~\cite{Pin, Oli} and the physical meaning of the estimation is not trivial. Here, we apply the speed limit inequality (\ref{main-q}) and estimate the bound of the equilibration time for several types of Hamiltonians and thermalization phenomena.

Let $\rho_{\rm S}'$ be a steady state of the subsystem inside the large total system. Then, the Bures angle between the initial state and the steady state after relaxation is given by $\calL_{\rm th}:= \arccos \left[   {\rm tr}\sqrt{ ( \rho_{\rm S}(0)^{1/2} \rho_{\rm S}' \rho_{\rm S}(0)^{1/2} )}\right]$. 
The steady state density matrix for the subsystem depends on the type of Hamiltonians. For instance, when the entire system is a static and uniform nonintegrable, the steady state of the subsystem is expected to be close to the canonical distribution of temperature determined by the initial state \cite{comment-thermalization}. In contrast, if the entire system is integrable, the steady state should be close to the generalized Gibbs ensemble \cite{dale}. Another type of thermalization is observed in the case of a nonintegrable system with a periodic driving field, which is sometimes referred to as Floquet thermalization. 
In this case, there are no conserved quantities, including energy. Owing to the energy injection via periodic driving, the periodic steady state of this system is a Gibbs state with infinite temperature. The reduced matrix for the subsystem should be $\rho_{\rm S}' = {\bm 1}/D$, where $D$ is the dimension of the Hilbert space of the subsystem, and in this case $\calL_{\rm th}$ is expressed as $\calL_{\rm th}= \arccos \left[   {\rm tr}( \rho_{\rm S}^{1/2}(0) /D^{1/2} )\right]$.

Regardless of the type of thermalization, the speed limit relation (\ref{main-q}) leads to the general lower bound of the subsystem equilibration time $\tau_{\rm eq}$ as
\begin{align}
  {\calL_{\rm th} \over \| \tilde{H}_S \| /\hbar  } \le {\calL_{\rm th} \over  \langle \delta E_{\rm S} \rangle_{\tau} /\hbar  }  \le
  \tau_{\rm eq} 
  \, .
\end{align}
The second term suggests that the average energy fluctuation is crucial for estimation of thermalization time. The first term can also be meaningful if the spectral norm is finite. 

\para{Derivation of \eqref{main-q}}
We present the proof of the speed limit for open quantum systems \eqref{main-q}.
We first consider the state change in short-time interval $\Di t$ at time $t$. We define the following unitary operators with an infinitesimal time duration $\Delta t$:
\balign{
U_1&:=e^{-i(\HS+H_{\rm SE})\Di t/\hbar} =e^{-i\tilde{H}_{\rm S} \Di t /\hbar}, \\
U_2&:=e^{-i H_{\rm E} \Di t/2\hbar}.
}
Here, note that $H_{\rm S}$, $H_{\rm SE}$ and $H_{\rm E}$ can be time-dependent, although we omit to write the explicit time-dependence for notational simplicity. We start with the following expression for the state of the total system at time $t +\Di t$:
\eq{
\rho_{\rm tot} (t+ \Di t)=U_2U_1U_2\rho_{\rm tot}(t)U_2^\dagger U_1^\dagger U_2^\dagger+o(\Di t ^2).
}
We now introduce
\balign{
\rho^{\prime}:=&U_2 \,\rho_{\rm tot}(t)\,U_2^\dagger , \\
{\rho}^{\prime\prime}
:=&U_1 U_2 \,\rho_{\rm tot} (t) \,U_2^{\dagger} U_1^{\dagger}
= e^{-i\tilde{H}_{\rm S}\Di t/\hbar}{\rho}^{\prime}e^{i\tilde{H}_{\rm S} \Di t/\hbar}, \label{rdbr}
}
where ${\rho}^{\prime\prime}$ is given by the time evolution from ${\rho}^{\prime}$ with the Hamiltonian $\tilde{H}_{\rm S}$. Below, we show that the quantum speed limit for the reduced density matrix $\rho_{\rm S}$ can be understood in terms of the density matrices $\rho^{\prime}$ and $\rho^{\prime\prime}$. We first note the cyclic property of the partial trace
\balign{
\rho_{\rm S}(t+ \Di t):=& \Tr_{\rm B}[\rho_{\rm tot } (t+ \Di t)]
=\Tr_{\rm E}\[ U_2\,{\rho}^{\prime\prime}\,U_2^\dagger \] + o(\Di t^2)   \nt \\
=& \Tr_{\rm E}{\rho}^{\prime\prime} + o(\Di t^2 ), \\
\rho_{\rm S}(t):=&\Tr_{\rm E}[\rho_{\rm tot} (t)]
=\Tr_{\rm E}\[ U_2^\dagger \rho_{\rm tot} (t) U_2 \] =\Tr_{\rm E}{\rho}^{\prime} .
}
We then note the monotonicity of the Bures angle $\calL$ for the contraction within the first order of $\Di t$:
\balign{
\calL(\rho_{\rm S}(t+ \Di t), \rho_{\rm S}(t))
&\leq \calL( {\rho}^{\prime\prime}, {\rho}^{\prime}).
}
From the relation (\ref{rdbr}) between $\rho^{\prime}$ and $\rho^{\prime\prime}$, the original MT speed limit \eqref{MT} leads to 
\balign{
\calL({\rho}^{\prime\prime}, {\rho}^{\prime})
\leq& \frac{1}{\hbar}\int_0^{\Di t} \delta {E}_S^{\prime} (t)  \, dt = \frac{\Di t}{\hbar}\delta {E}_S^{\prime}(t) 
}
up to the first order of $\Di t$. Here, the energy fluctuation $\delta {E}_S^{\prime} (t)$ is the energy fluctuation of $\tilde{H}_{\rm S}$ at time $t$ for ${\rho}^{\prime}$ defined as
\eq{
\delta {E}_{S}^{\prime} (t) :=\sqrt{\Tr [\tilde{H}_{\rm S} ^2 {\rho}^{\prime}]-\Tr [ \tilde{H}_{\rm S} {\rho}^{\prime}]^2 }\, .
}
Furthermore, we can equally treat $ \delta \tilde{E}_S^{\prime}(t)$ and $ \delta {E}_S^{\prime}(t)$ up to $O(\Di t)$:
\eq{
\frac{\Di t}{\hbar} \delta \tilde{E}_S^{\prime}(t)=\frac{\Di t}{\hbar}\delta E_S (t)+o(\Di t^{(3/2)} )\, , 
}
because the difference between $\rho_{\rm tot}(t)$ and ${\rho}^{\prime} =e^{-iH_{\rm E} \Di t/2\hbar} \rho_{\rm tot}(t)e^{i H_{\rm E} \Di t/2\hbar}$
is $o(\Di t)$. 
Finally, the triangle inequality of the Bures angle leads to
\balign{
\calL(\rho_{\rm S}(\tau), \rho_{\rm S}(0))\leq& \sum_{n=0}^{N-1} \calL(\rho_{\rm S}((n+1)\Di t), \rho_{\rm S}(n\Di t)) \nt \\
 \leq& \frac{\Di t}{\hbar}\sum_{n=0}^{N-1} \delta E_S(n\Di t)
}
with $\tau=N\Di t$.
By considering the $\Di t\to 0$ limit, we arrive at the desired result \eqref{main-q}.

\para{Summary}
We established a Mandelstam-Tamm type quantum speed limit inequality for open quantum systems with arbitrary coupling to general environments. Similar to the original Mandelstam-Tamm speed limit, the energy fluctuation still plays a pivotal role in this speed limit. Notably, the energy of the system and the interaction energy are relevant to this speed limit, and the energy of the environment is irrelevant. As potential applications, we discussed microscale quantum engines and the equilibration time in thermalization phenomena of isolated quantum systems. Relating the present speed limit to known results for Markovian dynamics can be a possible future scope of research.

\bigskip

\para{Acknowledgement}
We are grateful to Masaru Hongo for his questions that triggered this research.
NS was supported by JSPS Grants-in-Aid for Scientific Research Grant Number JP19K14615. 
KS was supported by JSPS Grants-in-Aid for Scientific Research (JP16H02211 and JP19H05603).

\clearpage

\pagestyle{empty}

\makeatletter
\long\def\@makecaption#1#2{{
\advance\leftskip1cm
\advance\rightskip1cm
\vskip\abovecaptionskip
\sbox\@tempboxa{#1: #2}%
\ifdim \wd\@tempboxa >\hsize
 #1: #2\par
\else
\global \@minipagefalse
\hb@xt@\hsize{\hfil\box\@tempboxa\hfil}%
\fi
\vskip\belowcaptionskip}}
\makeatother
\newcommand{\vo}{\upsilon}
\newcommand{\midskip}{\vspace{3pt}}

\setcounter{equation}{0}
\def\theequation{A.\arabic{equation}}

\begin{widetext}

\begin{center}
{\large \bf Supplemental Material for  \protect \\ 
  ``Speed limit for open systems coupled to general environments'' }\\
\vspace*{0.3cm}
Naoto Shiraishi$^{1}$ and Keiji Saito$^{2}$ \\
\vspace*{0.1cm}
$^{1}${\small \it Department of Physics, Gakushuin University, 1-5-1 Mejiro, Toshima-ku, Tokyo, Japan} \\
$^{2}${\small \it Department of Physics, Keio University, Yokohama 223-8522, Japan} 
\end{center}

\setcounter{equation}{0}
\renewcommand{\theequation}{S.\arabic{equation}}

\section{Derivation of Mandelstam-Tamm quantum speed limit}

\subsection{Case of pure state}
We first consider a situation that a pure state $\ket{\psi (0)}$ evolves to $\ket{\psi(\tau)}$ with a unitary evolution with duration $\tau$ generated by a time-dependent Hamiltonian $H(t)$. The case of mixed states is considered later.

We expand $\ket{\psi(t)}$ as $\ket{\psi (t)}=\sum_n a_n \ket{E_n}$, where $\ket{E _n}$ is the energy eigenstate of $H(t)$. 
We here drop the time dependence of $a_n$ for brevity.
Then the inner product between the state at times $t$ and $t+\Di t$ is calculated as
\balign{
\abs{\braket{\psi(t)|\psi(t+\Di t)}}
=&\abs{\sum_n \abs{a_n}^2 e^{-i E_n \Di t/\hbar}} \nt \\
=&\abs{\( 1-\frac{1}{2\hbar ^2}\sum_n \abs{a_n}^2 E_n^2 \Di t^2\)  -\frac{i}{\hbar}\sum_n \abs{a_n}^2 E_n \Di t +O(\Di t^3)} \nt \\
=&1-\frac{1}{2\hbar ^2}\sum_n \abs{a_n}^2 (E_n-\bar{E})^2 \Di t^2 +O(\Di t^3), \lb{pure-1}
}
where we defined the energy fluctuation as
\eq{
\bar{E}:=\sum_n \abs{a_n}^2 E_n.
}
We also denote the energy fluctuation by
\eqa{
\Di E^2:=\sum_n \abs{a_n}^2 (E_n-\bar{E})^2,
}{pure-2}
which is equivalent to the conventional definition $\Di E^2=\braket{\psi |H^2|\psi}-\braket{\psi|H|\psi}^2$.
Using an elemental inequality $\arccos x\leq \sqrt{2(1-x)}$ and the triangle inequality for the Bures angle $\calL(\rho, \rho')\leq \calL(\rho, \rho'')+\calL(\rho'', \rho')$, we find
\eq{
\calL(\ket{\psi(0)}, \ket{\psi(\tau)})\leq \sum_{n=0}^{N-1} \calL(\ket{\psi(t_n)}, \ket{\psi(t_{n+1})}) \leq \sum_{n=0}^{N-1} \sqrt{2(1-\abs{\braket{\psi(t_n)|\psi(t_{n+1})}})}
}
with $N\Di t=\tau$ and $t_n:=n\Di t$.
By combining Eqs.\eqref{pure-1} and \eqref{pure-2} and taking $\Di t\to 0$ limit, we arrive at the desired Mandelstam-Tamm inequality:
\eqa{
\calL(\ket{\psi (0)}, \ket{\psi(\tau)})\leq \frac{1}{\hbar}\int_0^\tau \sqrt{\Di E^2} dt.
}{MT-pure}

\subsection{Case of mixed state}

We now derive the Mandelstam-Tamm inequality for case of mixed states by using the obtained relation for pure states \eqref{MT-pure}.
The key ingredient of this extension is purification of mixed states to pure states by introducing an auxiliary system A.
For a given initial mixed state $\rho(0)$, there exists a proper auxiliary system and a pure state $\ket{\Psi(0)}$ on a composite system of the system S and the auxiliary system A satisfying $\Tr_{\rm A}[\ket{\Psi(0)}\bra{\Psi(0)}]=\rho(0)$.
An important fact is that if the state $\rho(0)$ evolves to $\rho(\tau)$ with the Hamiltonian $H(t)$, then $\ket{\Psi(0)}$ evolves to $\ket{\Psi(\tau)}$ with the Hamiltonian $H(t)\otimes {1}$ satisfying $\Tr_{\rm A}[\ket{\Psi(\tau)}\bra{\Psi(\tau)}]=\rho(\tau)$.
We note that both the expectation energy and the energy fluctuation take the same value between the original state and the purified pure state:
\balign{
\Tr_{\rm S}[H(t)\rho(t)]&=\Tr_{\rm SA}[(H(t)\otimes 1)\ket{\Psi(t)}\bra{\Psi(t)}] \\
\Tr_{\rm S}[H(t)^2 \rho(t)]&=\Tr_{\rm SA}[(H(t)\otimes 1)^2\ket{\Psi(t)}\bra{\Psi(t)}].
}
Then, the combination of the Mandelstam-Tamm relation for pure states \eqref{MT-pure} and the monotonicity of the Bures angle directly implies the desired Mandelstam-Tamm relation for mixed states:
\balign{
\calL(\rho(0),\rho(\tau))\leq \calL(\ket{\Psi(0)}, \ket{\Psi(\tau)}) \leq& \frac{1}{\hbar}\int_0^\tau \sqrt{\Tr_{\rm SA}[(H(t)\otimes 1)^2\ket{\Psi(t)}\bra{\Psi(t)}]-\Tr_{\rm SA}[(H(t)\otimes 1)\ket{\Psi(t)}\bra{\Psi(t)}]^2} dt \nt \\
 =& \frac{1}{\hbar}\int_0^\tau \sqrt{\Di E^2} dt,
}
where $\Di E^2$ is the energy fluctuation of the original system defined as $\Di E^2:=\Tr_{\rm S}[H(t)^2 \rho(t)]-\Tr_{\rm S}[H(t)\rho(t)]^2$.

\section{Derivation of the speed limit for classical closed systems}

We here derive the speed limit expression for the isolated classical system, which is first proved by Shanahan, {\it et al}.~\cite{Campoa}. 
Let $\rho (t)$ be a distribution function over the whole phase space at time $t$. 
The dynamics is given by the Liouville equation:
\begin{align}
  {d \over dt } \rho (t) &=  \left\{ H (t) \, , \rho (t) \right\}  =: -i\, \mathbb{L}_t\, \rho (t) \, . 
\end{align}
Note that this type of equation is also satisfied for $\sqrt{\rho (t)}$, i.e., ${(d /dt )\sqrt{\rho_t} } =  \left\{ H (t) \, , \sqrt{\rho (t)} \right\} $.

We define the Bures angle via the Bhattacharyya coefficient
\begin{align}
  {\cal L}_B (\rho (t) , \rho (0) ) &:= \arccos {\cal B} (t) \, , \\
  {\cal B} (t) &:=\int d\Gamma \sqrt{\rho (t)}\sqrt{\rho (0)} \, ,
\end{align}
where $\int d\Gamma$ stands for the integration over the whole phase space. From these definitions, it is straightforward to derive the following relation:
\begin{align}
  -  \sin {\cal L}_B {d {\cal L}_B \over dt} &=  \dot{\cal B}  \label{cb1} \\
                                             &= \int d\Gamma \sqrt{\rho (0)} \left\{ H (t) , \sqrt{\rho (t)} \right\} \, \\
               &=\int d\Gamma (\sqrt{\rho (0)} - \sqrt{\rho (t)} ) \left\{ H (t) , \sqrt{\rho (t)} \right\} \, , 
\end{align}
where we used an identity $\int d\Gamma  \sqrt{\rho (t)} \left\{ H (t) , \sqrt{\rho (t)} \right\} =\int d\Gamma \frac12 \frac{d}{dt}(\sqrt{\rho(t)})^2=0$. Using the Schwartz inequality, one can bound $\dot{\cal B}$
from above:
\begin{align}
  \dot{\cal B} &\le
                   \left[\int d\Gamma  \left(\sqrt{\rho (0)} - \sqrt{\rho (t)} \right)^2 \right]^{1/2}
  \left[ \int d\Gamma  \left( \left\{ H (t) , \sqrt{\rho (t)} \right\} \right)^2 \right]^{1/2} \, \\
                 &=  \left[ 2 (1  - {\cal B}(t) ) \right]^{1/2}   \left[ \int d\Gamma  \left( \left\{ H (t) , \sqrt{\rho (t)} \right\} \right)^2 \right]^{1/2}  \, \\
  &= 2 \sin ( {\cal L}_B /2 )   \left[ \int d\Gamma  \left( \left\{ H (t) , \sqrt{\rho (t)} \right\} \right)^2 \right]^{1/2}  \, ,
\end{align}
where we note that $0\le {\cal L}_B \le \pi $. 
Combining (\ref{cb1}) and this inequality, we have
\begin{align}
  2 {d \over dt} \sin ({\cal L}_B/2)  \le \left[ \int d\Gamma  \left( \left\{ H (t) , \sqrt{\rho (t)} \right\} \right)^2 \right]^{1/2}  \, ,
\end{align}
which directly leads to
\begin{align}
 2 \sin \left[ {\cal L}_B (\rho (\tau) , \rho (0) ) /2 \right]&= 2 \int_0^{\tau} dt\,  {d \over dt }\sin ({\cal L}_B (\rho (t) , \rho (0)) /2 )  \, \\
  &\le \int_0 ^{\tau} dt \left[ \int d\Gamma  \left( \left\{ H (t) , \sqrt{\rho (t)} \right\} \right)^2 \right]^{1/2}  \, .
\end{align}
With noting the inequality $(1/\pi) {\cal L}_B \le \sin ({\cal L}_B/2)$, we arrive at the following relation: 
\begin{align}
 {  {\cal L}_B (\rho (\tau), \rho (0) ) \over 
  {\pi \over 2 \tau} \int_{0}^{\tau} \left[ \int d\Gamma  \left( \left\{ H (t) , \sqrt{\rho (t)} \right\} \right)^2 \right]^{1/2}  }
  \le \tau \, . 
                 \label{class1}
\end{align}

\section{Derivation of the relation (\ref{main-c})}

We now provide the proof of \eqref{main-c}.
We follow a similar procedure to the quantum case. We divide the Liouville operator into three contributions, i.e.
\begin{align}
  \mathbb{L}_{\rm tot}(t) &=  \mathbb{L}_{\rm S} (t) + \mathbb{L}_{\rm SE} (t) + \mathbb{L}_{\rm E} (t) \, ,
\end{align}
where the subscripts ${\rm S}$, ${\rm SE}$, and ${\rm E}$ mean that the contribution from the system, coupling part between the system and environment, and the environment. Note the following identity
\begin{align}
   \int d\Gamma_{\rm E}  e^{- i \mathbb{L}_{\rm E} (t) dt} A (\Gamma ) &=   \int d\Gamma_{\rm E}  A (\Gamma ) \, , 
\end{align}  
where $\Gamma_{\rm E}$ is the phase space for the environment, and the function $A$ is arbitrary function dependent on distribution function.

We note the relations
\begin{align}
  d {\cal L}_B (\rho' , \rho'')  & \le dt {\pi \over 2 } \left[ \int d\Gamma  \left( \left\{ H_{\rm S} (t) + H_{\rm SE} (t) , \sqrt{\rho_{\rm tot} (t)} \right\} \right)^2 \right]^{1/2}  \, , \\
  \rho' &= e^{ -i dt \,( \mathbb{L}_{\rm S} (t) + \mathbb{L}_{\rm SE} (t) )  } e^{-i dt \, \mathbb{L}_{\rm E}(t)/2}\rho_{\rm tot} (t) \, , \\
                                  \rho '' &= e^{-i dt \, \mathbb{L}_{\rm E} (t)/2} \rho_{\rm tot} (t) \, .         
\end{align}
Subsequently, we note that the expression of the reduced distribution function
\begin{align}
  \rho_{\rm S} (t) &= \int \Gamma_{\rm E} \, \rho_{\rm tot}(t) = \int \Gamma_{\rm E}\,  e^{-i dt \, \mathbb{L}_{\rm E} (t)/2} \rho_{\rm tot} (t)= \int \Gamma_{\rm E} \, \rho'' \, , \\
  \rho_{\rm S}(t+dt) &= \int \Gamma_{\rm E} \,  \rho_{\rm tot} (t+dt) \\ &
                                                  = \int \Gamma_{\rm E} \, e^{-i dt \mathbb{L}_{\rm E} (t)/2} e^{ -i ( \mathbb{L}_{\rm S}(t) + \mathbb{L}_{\rm SE}(t) ) dt } e^{-i dt \mathbb{L}_{\rm E}(t)/2}
                  \rho_{\rm tot}(t) + \mathcal{O} (dt^2) \\
               &= \int \Gamma_{\rm E} \,  e^{ -i ( \mathbb{L}_{\rm S}(t) + \mathbb{L}_{\rm SE}(t) ) dt } e^{-i dt \mathbb{L}_{\rm E}(t)/2}
                 \rho_{\rm tot}(t) + \mathcal{O} (dt^2)    \\ &= \int \Gamma_{\rm E} \,  \rho' ~ + \mathcal{O} (dt^2) \, . 
\end{align}
Through the contraction, the following relation is satisfied up to $\mathcal{O} (dt)$:
\begin{align}
   {\cal L}_B (\rho_{\rm S} (t+dt) , \rho_{\rm S} (t) ) \le  {\cal L}_B  (\rho' , \rho'')  \, .
\end{align}
Then we have
\begin{align}
  {\cal L}_B (\rho_{\rm S}(t+dt) , \rho_{\rm S}(t) )
  \le dt {\pi \over 2 }
  \left[ \int d \Gamma  \left( \left\{ H_{\rm S}(t) + H_{\rm SE}(t) , \sqrt{\rho_{\rm tot} (t)} \right\} \right)^2 \right]^{1/2}  \, .
\end{align}
By integrating this from $0$ to $\tau$, the desired result (\ref{main-c}) is obtained.

\end{widetext}


\end{document}